\documentclass[aps,prl,twocolumn,superscriptaddress,showpacs]{revtex4-1}

\usepackage{graphicx}
\usepackage{amsmath,amssymb}

\usepackage{bm}
\usepackage[T1]{fontenc} 
\usepackage{lmodern}
\usepackage{color}

\setcitestyle{super,open={},close={}}

\graphicspath{{./Figures/}}

\begin{document}

\title{Intelligence of agents produces a structural phase transition in collective behaviour}

\author{Hannes Hornischer}
\author{Stephan Herminghaus}
\author{Marco G. Mazza}
\affiliation{Max Planck Institute for Dynamics and Self-Organization (MPIDS), Am Fa{\ss}berg 17, 37077 G\"{o}ttingen, Germany}

\date{\today}

\begin{abstract}
Living organisms process information to interact and adapt to their changing environment with the goal of finding food, mates or averting hazards. The structure of their niche has profound repercussions by both selecting their internal architecture and also inducing adaptive responses to environmental cues and stimuli. Adaptive, collective behaviour underpinned by specialized optimization strategies is ubiquitously found  in the natural world. This exceptional success originates from the processes of fitness and selection. Here we prove that a universal physical mechanism of a nonequilibrium transition underlies the collective organization of information-processing organisms. As cognitive agents build and update an internal, cognitive representation of the causal structure of their environment, 
complex patterns  emerge in the system, where the onset of pattern formation relates to the spatial overlap of cognitive maps. 
Studying the exchange of information among the agents reveals a continuous, order-disorder transition. 
As a result of the spontaneous breaking of translational symmetry, a Goldstone mode emerges, which points at a collective mechanism of information transfer among cognitive organisms. 
Taken together, the characteristics of this phase transition consolidate different results in cognitive and biological sciences in a universal manner. 
These finding are generally applicable to the design of artificial intelligent swarm systems that do not rely on  centralized control schemes.
\end{abstract}
   
\maketitle

\begin{figure}
  \includegraphics[width=0.4\textwidth]{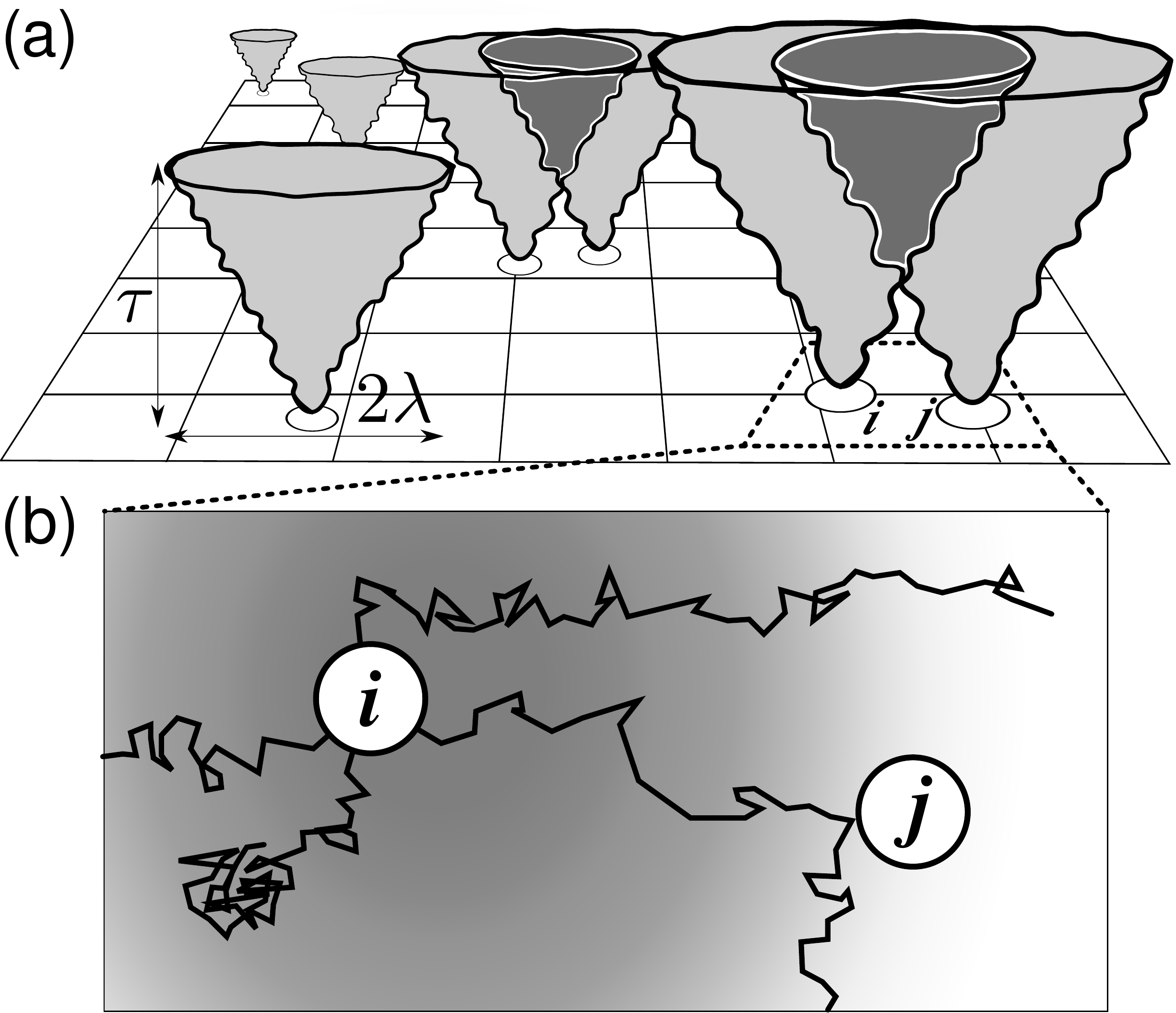}
  \caption{Schematic representation of a system of cognitive agents and their cognitive maps. (a) Starting from the initial condition in configuration space the agents (empty circles) create a cognitive map of their surroundings by means of hypothetical sampling trajectories of duration $\tau$. The cones show a $(2+1)$-dimensional envelop of the trajectories, which represent the cognitive maps, of average radius $\lambda$. When the cones overlap, the agents sense each other and a force arises. (b) Shown here are four sampling trajectories emanating from cognitive agent $i$. The grey gradient represents the projection of the cognitive map onto the plane. As one of the trajectories impacts with agent $j$, agent $i$ is forced to modify its trajectory, thus responding to its cognitive representation of the environment.  Cognitive competence is the tendency to maximize the options left after one agent enters the space of the other, and to avoid in the most efficient way the overlapping regions.}
  \label{fig:sketch_cones}
\end{figure}   

Living organisms respond to environmental cues and stimuli in order to locate resources, avoid threats, or express social behaviour~\cite{schultzScience1997,vergassolaNat2007,torneyPNAS2009}. The navigation of the environment requires a feedback loop of information processing, inference, and active response that optimizes the sensory inputs~\cite{vergassolaNat2007,torneyPNAS2009}, and that, in time, produces evolutionary survival strategies. In this view, living organisms constantly incorporate the spatio-temporal information of their dynamic environment into an internal representation. We call this internal representation of the geometry, topography, external entities, and other aspects of the environment a cognitive map~\cite{realScience1991,kamilNat1997,menzelPNAS2005,couzinNat2007,normandAnimalBehav2009}.
The emergence of adaptive organization  in the collective behaviour of cognitive agents naturally prompts the question whether we can find generic features which are characteristic of the collective dynamics of cognitive agents. In other words, does the cognitive competence of the agents result in characteristic features of their collective behaviour, and if so, what are these features?

In order to investigate this fundamental question, we need to choose a quantitative description of cognitive competence which allows to be implemented into a model of collective behaviour.  It is clear that there are many phenomenological aspects which distinguish an agent with some intelligence from a simple active matter agent, like a self-propelling particle. 
Internal representation is a highly complex cognitive process that can be realized through memory, pattern recognition, and maximization of the information about the environment. 
 
\begin{figure*}
\includegraphics[width=0.7\textwidth]{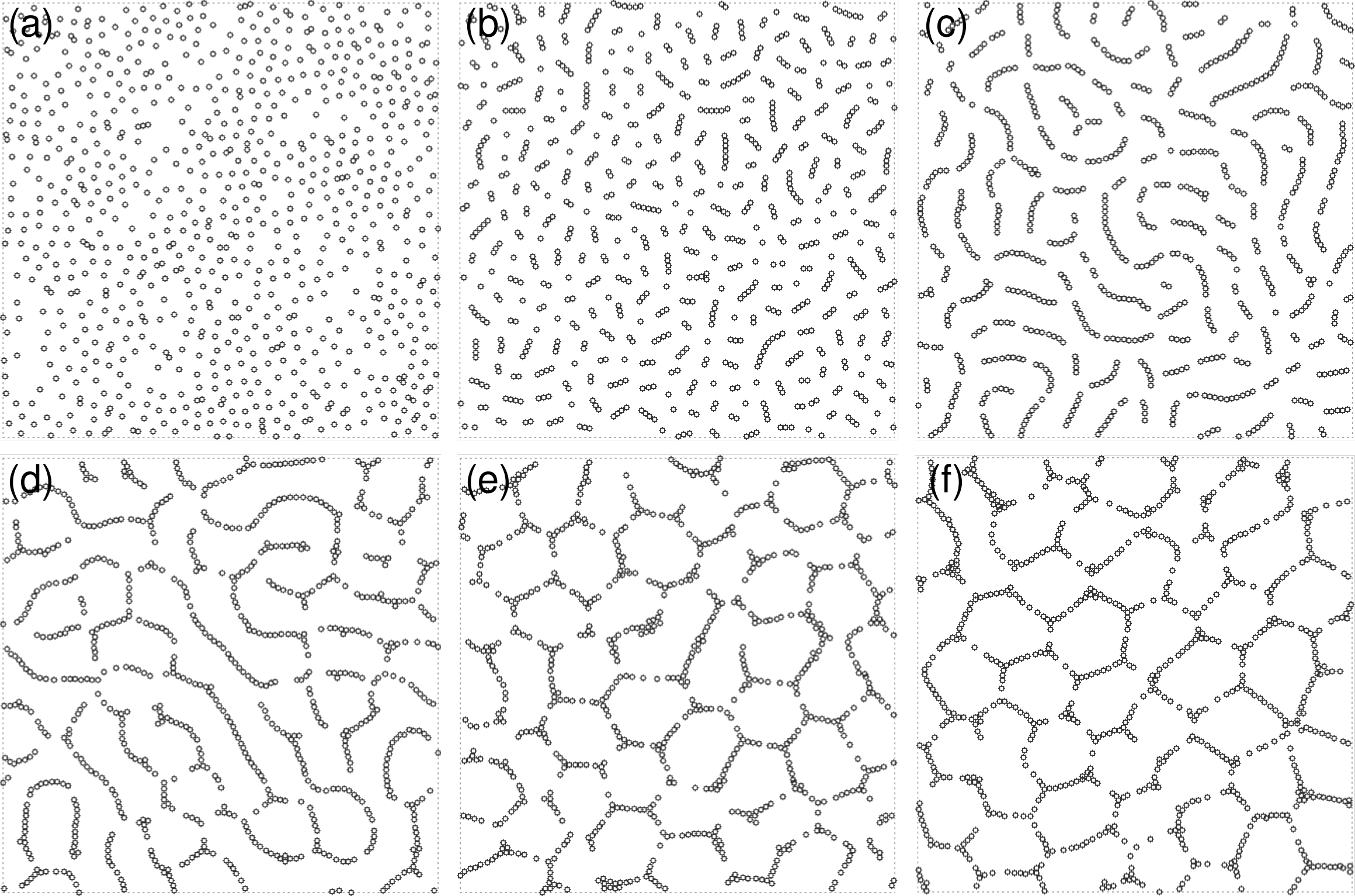}
\caption{Snapshots of a two-dimensional system system of $N=800$ ($L=80\sigma$) cognitive agents in steady-state configurations. As the size of the cognitive map $\lambda$ increases, complex structures emerge. (a) At $\lambda=0.8\sigma$ the agents are randomly distributed. (b) At $\lambda=3.2\sigma$ agents start forming short linear chains. (c) At $\lambda=5.6\sigma$ the chains outline a labyrinthine pattern. (d) At $\lambda=7.6\sigma$ the labyrinthine pattern continuously changes into a cellular structure.  At  $\lambda=10\sigma$ (e) and  $\lambda=10.8\sigma$ (f) the cellular structure is well formed.
Thus, cognitive agents capable of `intelligent' response to their environment and maximizing their cognitive map information exhibit a clear phase transition from disorder (at small $\lambda$) to order (at large $\lambda$).}
\label{fig:snapshots}
\end{figure*}

Cognitive maps are a prerequisite of intelligent behaviour.   Consider, for instance, a simple analogy to a chess player. It is the player's internal cognitive map that allows to contemplate the possible moves and the ramifications of their consequences, i.e., hypothetical trajectories. At the same time, the player's strategy can be cast into a simple, universal form: the maximization of future options to move without losing the king. The player who has no such option left is called checkmate and loses. For the present study, we identify cognitive competence with the ability of the agent to maximize the number of hypothetical moves within the agent's cognitive map of the environment.

Given that  an agent creates a representation of its environment within a certain distance, 
it is straightforward to quantify such representation for modelling.
We define a cognitive map as the ensemble of random hypothetical trajectories, $\{\Gamma_\tau(t)\}$, each of total duration $\tau$, that the agent may traverse to explore its environment. We explicitly indicate the dependence on time $t$ because the cognitive map is dynamically updated as information is acquired.
Starting from its initial position, $\textbf{x}_0$, the area probed by the agent, and thus the size of its cognitive map, is directly proportional to $\tau$. The number of options necessary to traverse such trajectories, $P(\Gamma_\tau(t)|\mathbf{x}_0)$, can be expressed as an information entropy of the cognitive map, $\mathcal{S}(\mathbf{X},\tau) =-k_\mathrm{B}\! \int\! P(\Gamma_\tau(t)|\mathbf{x}_0)\ln P(\Gamma_\tau(t)|\mathbf{x}_0) \,\mathcal{D}\Gamma_\tau(t)$, which is expressed as a path integral over the cognitive map $\{\Gamma_\tau(t)\}$ (see Methods and \cite{MartyushevPR2006,wissnerPRL2013}), and where  $k_\mathrm{B}$ is Boltzmann's constant to give dimensions of entropy. The symbol $\mathbf{X}$ represents the characteristics of the environment and of other agents.

Maximization of $\mathcal{S}$ can then be represented by a force acting on the agent of the form $\mathbf{F}(\mathbf{X},\tau) = \theta \nabla_{\mathbf{X}} \mathcal{S}(\mathbf{X},\tau)$. The coupling parameter $\theta$ (with dimensions of temperature)
 represents how high is the cognitive competence of the agent, that is, how strongly the agent responds to the environment (see Methods). In order to maximize the information an agent needs to take into account other agents. This corresponds to following the gradient of $\mathcal{S}$.
Similar expressions have been used in cosmology~\cite{boussoPRL2006,boussoPRD2007}, biological infotaxis~\cite{vergassolaNat2007}, sensorimotor systems~\cite{klyubinPLOS2008}, and control theory~\cite{wissnerPRL2013}.  
Importantly, maximization of information has been found as characterizing human cognition~\cite{guevaraPRE2016}.
Figure~\ref{fig:sketch_cones} shows a sketch of a few agents which interact with each other and with the environment. Each agent explores the available configuration space and acquires information about its structure, and in so doing builds its cognitive map, and optimizes its behaviour through responding to the surrounding. 
In the overlap regions of the forward cones the agents have a probability to collide, and because of this the effective force $\mathbf{F}(\mathbf{X},\tau)$ arises.
The overlap regions and the corresponding effective forces  appear when the distance between any two agents becomes shorter than the average linear length, $\lambda(\tau)$, of the agents' hypothetical trajectories, which relates to the size of the cognitive map.

Our definition of information entropy $\mathcal{S}$ satisfies the following criteria.  First, $\mathcal{S}$ is based on the information content of the system because agents retrieve and process information about the presence of other agents. Second, it is universal, as it does not require any specific goal or strategy, such as rules for taxis of bacteria in chemo-attractant concentration fields. Third, it obeys the laws of information theory and information processing, essential to build cognitive maps. Fourth, it obeys causality because the current state of the cognitive map influences the agent's future dynamics.

\begin{figure}
  \includegraphics[width=0.5\textwidth]{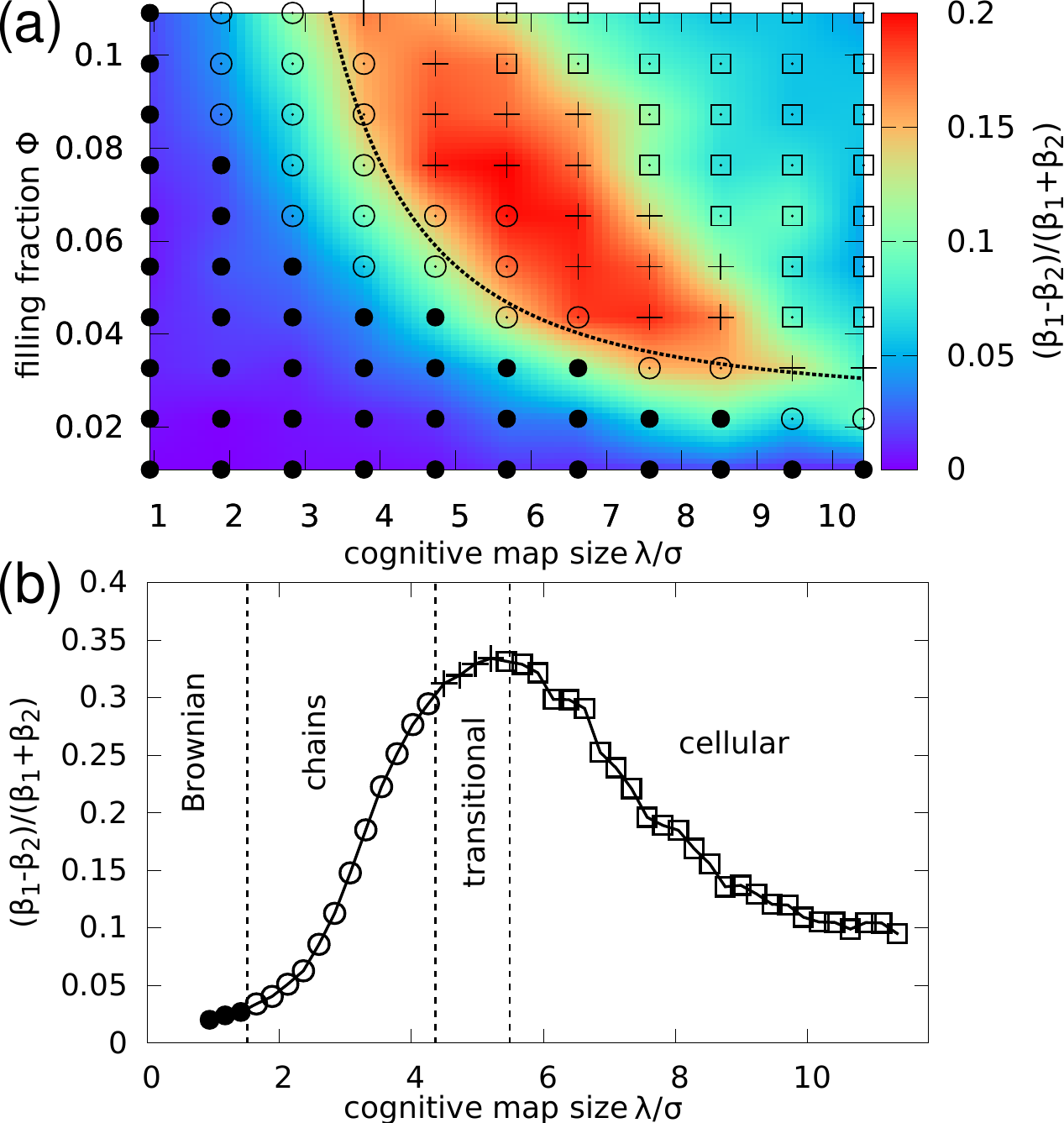}
  \caption{Nonequilibrium phase diagram of the cognitive agent system. (a) Symbols represent the steady-state configuration for up to $N=500$ agents: isolated agents ($\bullet$), short chains ($\circ$), labyrinthine pattern ($+$), cellular pattern ($\Box$). Heat map of the anisotropy $\alpha\equiv\frac{\beta_1-\beta_2}{\beta_1+\beta_2}$ of the $W_2^{1,1}$ Minkowski tensor. The dotted line represents the $\phi^{-1/2}$ scaling of the transition values of $\lambda$ (note that an offset is added because at very low filling fraction no transition is expected as there is now significant overlap of the cognitive maps). (b) Dependence of the anisotropy $\alpha\equiv\frac{\beta_1-\beta_2}{\beta_1+\beta_2}$ on the cognitive map size $\lambda$ for a system at $\phi= 0.1$ ($N=800$). Symbols have the same meaning as in panel (a). Cognitive agents exhibit an order-disorder phase transition to a cellular pattern as the mean inter-agent distance becomes comparable to the cognitive map size $\lambda$.}
  \label{fig:phasemap}
\end{figure}

We carried out simulations of $N$ agents in a two-dimensional, continuous system of size $L\times L$, where agents interact with each other via the force $\mathbf{F}$ and with hard-core repulsion when their distance is less than the agent diameter $\sigma$. Figure~\ref{fig:snapshots} shows the steady-state configurations of the system as the size $\lambda$ of the map increases. At low values of cognitive map size $\lambda$ with respect to the inter-agent separation (Fig.~\ref{fig:snapshots}(a)) most agents are isolated and randomly distributed throughout the system. As $\lambda$ increases, we observe the spontaneous formation of short linear chains of agents (Fig.~\ref{fig:snapshots}(b)). At $\lambda=5.6\sigma$ the chains grow longer and outline a labyrinthine pattern in the system (Fig.~\ref{fig:snapshots}(c)). Upon further increase of $\lambda$, the pattern continuously morphs into a cellular structure (Fig.~\ref{fig:snapshots}(d-e)), which we find well developed at $\lambda=10.8\sigma$ (Fig.~\ref{fig:snapshots}(f)). A similar sequence of patterns can be observed when we vary the filling fraction, $\phi\equiv N\pi\sigma^2/L^2$. The phase diagram of the system is shown in Figure~\ref{fig:phasemap}(a).  The transition line from short chains to more complex pattern is well fitted by a relation $\phi\sim \lambda^{-2}$ which suggests that the transition is triggered as the mean inter-agent distance is comparable to the cognitive map size $\lambda$.

In order to analyse the complex morphology of the patterns, we employ the anisotropy parameter $\alpha\equiv\frac{\beta_1-\beta_2}{\beta_1+\beta_2}$, defined in terms of the eigenvalues $\beta_1$ and $\beta_2$ of the 
Minkowski tensor~\cite{schroederturkJMicro2010,schroderturkNJP2013} (see Methods for more details). 
Figure~\ref{fig:phasemap}(a) shows as a heat map the association of the phase diagram with the anisotropy $\alpha$ of the configurational patterns.
At fixed filling fraction $\phi$, the system exhibits the largest anisotropy
$\alpha$ when linear chains start to connect with each other for intermediate values of the size $\lambda$ of the cognitive map.  
In contrast, at low $\lambda$, where agents are isolated, the system is trivially isotropic. At large values of $\lambda$, where cognitive maps significantly overlap and cellular patterns emerge (Fig.~\ref{fig:snapshots}(f)), the associated anisotropy decreases to values that are however larger than in the case of low $\lambda$. This indicates that the system gains again isotropy on the larger scale of the cells.  Figure~\ref{fig:phasemap}(b) shows the anisotropy $\alpha$ of the pattern.  It exhibits a sharp maximum at $\lambda\approx 5.5$ where the linear chains are most pronounced and  the system is at the threshold of forming the labyrinthine patterns. 

The continuous dependence of $\alpha$ on $\lambda$ provides evidence of a continuous transition.  It proves instructive to consider the mutual flow of information that ensues from the overlap of the cognitive maps and, in turn,  
dynamically modifies the cognitive maps of the agents. This information flow can be quantified via the notion of mutual information~ \cite{schreiberPRL2000,wicksPRE2007,MelzerPRE2014}.  The mutual information for two sequences $(a_i)$ and $(b_j)$ is given by $\sum_{i,j} {P}(a_i,b_j) \log_2\frac{{P}(a_i,b_j)}{{P}(a_i){P}(b_j)}$, where ${P}(a_i)$ represents the probability of occurrence of of the $a_i$ value, while ${P}(a_i,b_j)$ is the joint probability for the values $a_i$ and $b_j$. Because we are interested in isolating the causal interaction between agents that underpins the update of the cognitive maps, we consider the positions  $(x_i(t),y_i(t))$ of the $i$th agent at time $t$ and compute $\vec{M}_{ij}\equiv(M^x_{ij},M^y_{ij})$, where $M^x_{ij}={P}(x_i,x_j) \log_2\frac{{P}(x_i,x_j)}{{P}(x_i){P}(x_j)}$, and $M^y_{ij}={P}(y_i,y_j) \log_2\frac{{P}(y_i,y_j)}{{P}(y_i){P}(y_j)}$. The total mutual information is then 
\begin{equation}\label{eq:totMI}
\mathcal{M}=\frac{1}{N_p}\sum_{i,j\in \mathcal{N}(i)} |\vec{M}_{ij}|\,, 
\end{equation}
where $\mathcal{N}(i)$ is the set of agents whose distance from $i$ is smaller than $4\sigma$, and $N_p\equiv \sum_i|\mathcal{N}(i)|$ is the total number of pairs $(i,j)$ included in the sum in Eq.~\eqref{eq:totMI}. We note that the hypothetical Brownian  trajectories are determined by a temperature $T$, and therefore a longer $\lambda$ relates to a higher $T$.

\begin{figure}
  \includegraphics[width=0.45\textwidth]{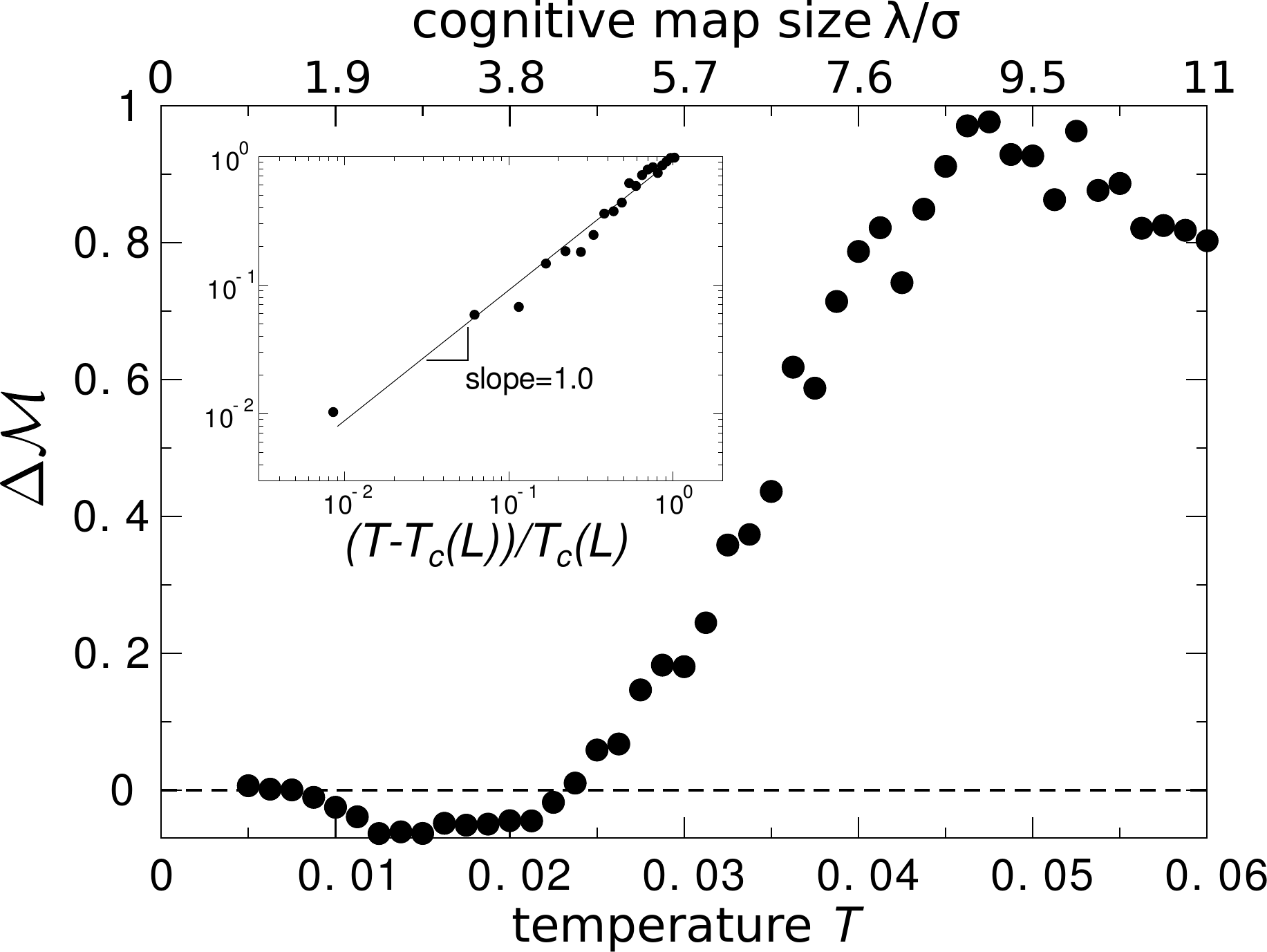}
\caption{Nonequilibrium transition in a cognitive-agent system. As the temperature $T$ associated with the hypothetical Brownian  trajectories increases, the mutual information among the cognitive agents increases once their maps overlap significantly. 
The main panel shows the dependence of the relative mutual information  $\Delta\mathcal{M}$ on $T$ for $L=80\sigma$  at fixed number density. The results are averaged over eight independent simulations for each data point. The inset shows the scaling of $\Delta\mathcal{M}$ in proximity of the critical temperature $T_c$. The cognitive agent system exhibits a continuous phase transition to complex patterns as the overlap of their cognitive maps grows and they maximize the information content of the map.}
  \label{fig:MI}
\end{figure}

\begin{figure}
  \includegraphics[width=0.4\textwidth]{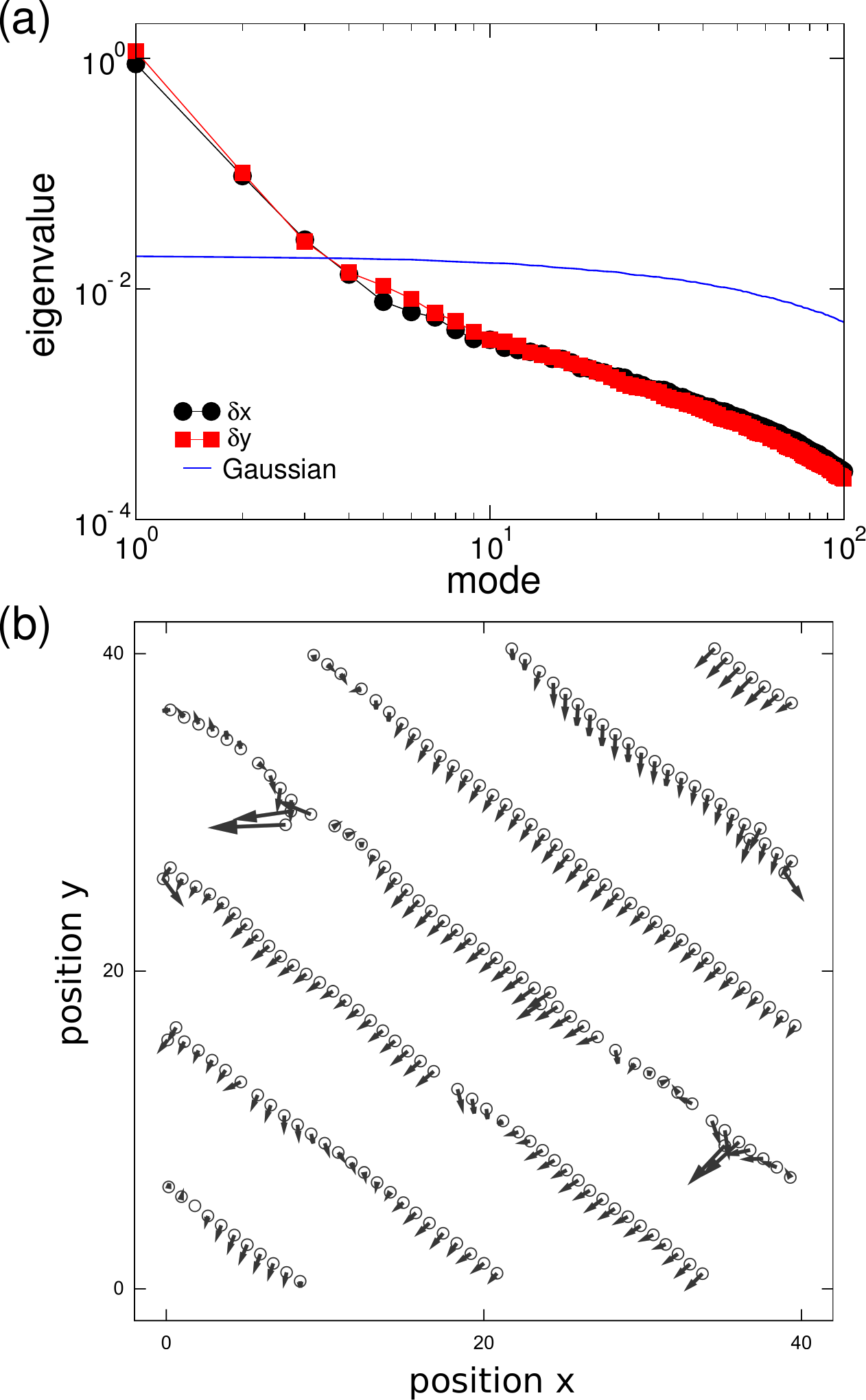}
\caption{Growth of correlated motion. (a) Spectrum of eigenvalues of the displacement covariance matrix (for a system with $\phi=0.1$ and $\lambda=8\sigma$)
and comparison with a model of Gaussian-distributed random displacements. The largest eigenvalue mode corresponds to a Goldstone mode propagating through the system. (b) We show the collective fluctuation in the agents' positions due to a Goldstone mode propagating through the system when elongated linear chain patterns emerge. As complex patterns emerge and Galilean invariance is spontaneously broken, collective fluctuations materialize through the propagation of a Goldstone mode.}
  \label{fig:Eigenmodes}
\end{figure}

As a measure of the impact of the cognitive maps onto the trajectories, we consider the relative mutual information $\Delta\mathcal{M}=\mathcal{M}-\mathcal{M}_\mathrm{Br}$,
where $\mathcal{M}_\mathrm{Br}$ is calculated following Eq.~\eqref{eq:totMI} for a system of Brownian agents characterized by the same $T$ as in the cognitive agent system.
 Figure~\ref{fig:MI}(a) shows the dependence of the relative mutual information $\Delta\mathcal{M}$ on temperature $T$. At very low $T$, the cognitive agent system exhibits a mutual information which is within scattering consistent with a collection of Brownian agents~\footnote{The shallow minimum around $T\approx 0.015$ can be ascribed to the formation of  dimers and trimers which have a reduced mobility. This effect will be disregarded here}. 
As the temperature increases beyond a critical value $T_c\approx 0.023$, $\Delta\mathcal{M}$  increases steadily, while the system develops labyrinthine and cellular patterns.

Taking $T_c$ as the critical temperature of the order-disorder transition reveals a region of critical scaling of $\Delta\mathcal{M}$ with rescaled temperature. The critical exponent is unity  within data scattering (see inset of Fig.~\ref{fig:MI}). At $T\approx 0.05$, $\Delta\mathcal{M}$ reaches a maximum, and for even larger $T$ decreases slightly to a plateau value. 

In order to further investigate the analogy with continuous phase transitions, we consider the spontaneous breaking of translational symmetry (Galilean invariance)  associated with the emergence of the patterns for $T>T_c$. In this case we expect a Goldstone mode to emerge, which corresponds to `massless' excitations propagating in the system at infinite correlation length. 
 
 As we need to analyse the dynamic response of the agents we consider the displacement covariance matrix~\cite{henkesSoftMatt2012} $C_{ij}\equiv \langle \delta\mathbf{r}_i(t)\cdot\delta\mathbf{r}_j(t) \rangle$, where $\delta\mathbf{r}_i(t)\equiv\mathbf{r}_i(t)-\langle \mathbf{r}_i \rangle$. Figure~\ref{fig:Eigenmodes} shows the eigenvalues of $C_{ij}$ for $\delta x$ and $\delta y$, and the comparison with a random matrix model of Gaussian-distributed displacements. The first mode is considerably above the random Gaussian model, and corresponds to a large wavelength mode propagating through the system (shown in the inset of Fig.~\ref{fig:Eigenmodes}). This is the Goldstone mode associated to the order-disorder transition and corresponds to excitations of the ordered state.

In conclusion, we have found the first nonequilibrium phase transition in a system of cognitive agents that dynamically interact with their environment and respond to it with cognitive competence by maximizing the information content of their cognitive maps~\cite{guevaraPRE2016}.
The transition from isolated Brownian-like particles to complex patterns is characterized by different degree of overlap of the cognitive maps. 
%
The continuous change of $\Delta\mathcal{M}$ as the system develops complex patterns, together with the  change of the anisotropy parameter $\frac{\beta_1-\beta_2}{\beta_1+\beta_2}$  point at a continuous transition in cognitive-agent systems. We have found the existence of critical scaling at the onset of the transition, where the cognitive maps overlap significantly. 
 We have identified a Goldstone mode propagating through the system that is generated in the spontaneous symmetry breaking of the order-disorder transition as complex patterns emerge. 
Our results are relevant to artificial systems like autonomous micro-robots, and swarm robotic~\cite{sahinIntWorkSwarm2004,brambillaSwarmInt2013}
systems explicitly designed to autonomously mimic the collective behaviour of living organisms.

\noindent
{\small
\textbf{Methods}\\
\noindent
\textbf{Construction of the cognitive map}.
The calculation of $P(\Gamma_\tau(t)|\mathbf{x}_0)$ is performed by generating hypothetical sampling trajectories, each of which represents a virtual evolution during the time $[0,\tau]$ of the agent with constraints fixed at the present configuration and not depending on time (they correspond to D'Alembert virtual displacements). The virtual trajectories are generated using Langevin dynamics $m\mathbf{\dot{v}}=-\gamma \mathbf{v}+\bm{\xi}(t)+\mathbf{h}(\mathbf{r})$, 
where $\mathbf{v}$ is the velocity of the agent, $m$ its mass, $\gamma$ the viscous drag, $h(\mathbf{r})$ represents the holonomic constraints and the other agents, and $\bm{\xi}(t)$ is a random noise with zero mean and $\langle\xi_i(t)\xi_j(t')\rangle=2\gamma k_\mathrm{B}T\delta_{ij}\delta(t-t')$.\\
\noindent
\textbf{Simulations}.
For every agent the force due to the overlapping cognitive maps 
$\mathbf{F}(\mathbf{X},\tau)= \frac{2\theta}{TN_\Omega}  \langle\sum_{n=1}^{N_\Omega} \mathbf{f}_{n}(0)  \ln(\frac{\Omega_n}{\langle\Omega_{n}\rangle }) \rangle$
is calculated by independently sampling the phase-space volume $\Omega$ via $N_\Omega$ trajectories. The effective temperatures $T$ and $\theta$ and the random force $\mathbf{f}_n(0)$ determine the first step of the hypothetical sampling trajectory. 
The ratio of the temperatures ${\theta}/{T}$ determines the magnitude of the causal entropic force. The parameter $T$ is a  measure for the random noise and is proportional to the average linear size $\lambda$
of a hypothetical sampling trajectory.  The phase-space volume of every trajectory is defined as 
$\Omega_i^{-1}=N_\Omega P(\Gamma_\tau(t)|\mathbf{x}_0)$, 
and we approximate it through the radius of gyration $R$ of all positions of the sampling trajectory relative to their mean position $\Omega_i \approx R^2 = \frac{1}{K}\sum_{k=1}^{K}(\mathbf{r}_k-\bar{\mathbf{r}})$ and then applied to it according to the equation of motion $m\mathbf{\dot{v}}=-\gamma \mathbf{v}+\mathbf{F}(t)+\mathbf{h}(\mathbf{r})$. All agents have the diameter $\sigma$. Any interaction of a sampling trajectory with another agent is hard core, whereas actual agents interact with each other through a repulsive linear spring when $|\mathbf{r}_i-\mathbf{r}_j|<\sigma$. The agents are initially placed randomly with a uniform distribution within the system and without any overlap of the agents' hard cores.
The system size is fixed at $L=80\sigma$.\\
\noindent
\textbf{Measure of anisotropy}.
The Minkowski tensor $\mathbf{W}_{2}^{1,1}(C)\equiv\frac{1}{2}\int_{\partial C} \mathbf{r}\odot\mathbf{n}\, G_{2}\,\textrm{d}r$ provides a measure of anisotropic morphologies~\cite{schroederturkJMicro2010,schroderturkNJP2013}. It is a second rank symmetric tensor, where $G_2=(\kappa_1+\kappa_2)/2$ is the local curvature, $\mathbf{r}$ the position vector, $\mathbf{n}$ the normal vector to the surface $\partial C$ of a body, and $(a \odot b)_{ij}\equiv (a_ib_j+a_jb_i)/2$ is the symmetric tensor product of vectors $\mathbf{a}$ and $\mathbf{b}$. The 
anisotropy parameter $\alpha\equiv\frac{\beta_1-\beta_2}{\beta_1+\beta_2}$, where $\beta_1$ and $\beta_2$ are the   largest and smallest eigenvalues of $\mathbf{W}_{2}^{1,1}$, respectively, gives a measure of anisotropy of the pattern.

\noindent
\textbf{Acknowledgments}\\
We gratefully acknowledge discussions with Giovanni Ciccotti, Mirko Lukovic, Paolo Malgaretti, Jan Nagler,  Agostina Palmigiano, and Andr\'e Schella.

}


\end{document}